\documentclass[aps,prl,twocolumn,superscriptaddress, showpacs]{revtex4-1}

\usepackage{graphicx}
\usepackage{bm}
\usepackage{amssymb}
\usepackage{pifont}

\begin{document}


\title{Direct measurement of general quantum states using
weak measurement}

\author{Jeff S. Lundeen}

\email{jeff.lundeen@nrc-cnrc.gc.ca}

\author{Charles Bamber}
\affiliation{Institute for National Measurement Standards, National Research Council,
1200 Montreal Rd., Ottawa, ON, K1A 0R6}

\begin{abstract}
Recent work {[}J.S. Lundeen et al. Nature, \textbf{474}, 188 (2011){]}
directly measured the wavefunction by weakly measuring a variable
followed by a normal (i.e. `strong') measurement of the complementary
variable. We generalize this method to mixed states by considering
the weak measurement of various products of these observables, thereby
providing the density matrix an operational definition in terms of
a procedure for its direct measurement. The method only requires measurements
in two bases and can be performed `in situ', determining the quantum
state without destroying it. 
\end{abstract}


\maketitle

The wavefunction $\Psi$ is at the heart of quantum mechanics, yet
its nature has been debated since its inception. It is typically relegated
to being a calculational device for predicting measurement outcomes.
Recently, Lundeen et al. proposed a simple and general operational
definition of the wavefunction based on a method for its direct measurement:
{}``it is the average result of a weak measurement of a variable
followed by a strong measurement of the complementary variable \cite{Lundeen2011, Lundeen2006}.''
By `direct' it is meant that a value proportional to the wavefunction
appears straight on the measurement apparatus itself without further
complicated calculations or fitting. The `wavefunction' referred to
here is a special case of a general quantum state, known as a `pure
state.' The general case is represented by the density operator $\rho$,
which can describe both pure and `mixed' states. The latter incorporates
both the effects of classical randomness (e.g., noise) and entanglement
with other systems (e.g., decoherence). The density operator plays
an important role in quantum statistics, quantum information, and
the study of decoherence. Because of its generality and because it
follows naturally from classical concepts of probability and measures,
some consider $\rho$, rather than $\Psi$, to be the fundamental
quantum state description. In this letter, we propose two methods
to directly measure general quantum states, one of which directly
gives the matrix elements of $\rho$.

The standard method for experimentally determining the density operator
is Quantum State Tomography \cite{Vogel1989,*Smithey1993,*Breitenbach1997,*White1999}.
In it, one makes a diverse set of measurements on an ensemble of identical
systems and then determines the quantum state that is most compatible
with the measurement results. An alternative is our direct measurement
method, which may have advantages over tomography, such as simplicity,
versatility, and directness. A quantitative comparison of measures
such as the signal to noise ratio, resolution, and fidelity, has not
been undertaken but some limitations of the direct method have been
identified in \cite{Haapasalo2011}. As compared to tomography, which
works with mixed states, the most significant limitation of the direct
measurement of the wavefunction is that it has only been shown to
work with pure states.

Previous works have developed direct methods to measure quasi-probability
distributions, such as the Wigner function \cite{Wigner1932}, Husimi
Q-function \cite{Husimi1940}, and the Glauber-Sudarshan P-function
\cite{Glauber1963,*Sudarshan1963}. These are position-momentum (i.e.
`phase-space') distributions that are equivalent to the density operator,
and have many, but not all, of the properties of a standard probability
distribution. The Wigner function can be directly measured by displacing
the system in phase space and then measuring the parity operator \cite{Banaszek1996,*Laiho2010}.
Equivalently, the integral of the interference between a pair of rotated
and displaced replicas of the system will give the Wigner function
\cite{Mukamel2003,*Smith2005}. The Husimi Q-function can be directly
measured by an eight-port homodyne apparatus or by projection on the
harmonic oscillator ground state \cite{Leonhardt1997,*Kanem2005}.
These phase-space distributions are created to be the closest quantum
analogs to a classical probability distribution. In this sense, they
are inherently amenable to direct measurement.

\textit{Weak measurement.}---We begin by considering what happens
to our method for directly measuring the wavefunction when the state
is not pure. At the heart the direct method is weak measurement \cite{Aharonov1988}.
Over the last decade, interest in weak measurement has grown as researchers
have realized its potential for interrogating quantum systems in a
coherent manner \cite{Aharonov2010,*Cho2011}. It has been used to
model and understand photonic phenomena in birefringent photonic crystals
\cite{Solli2004}, fiber networks \cite{Brunner2003,*Brunner2004},
cavity QED \cite{Wiseman2002}, and quantum tunneling \cite{Steinberg1995, Steinberg1995a}.
Weak measurement provides insight into a number of fundamental quantum
effects, including the role of the uncertainty principle in the double-slit
experiment \cite{Wiseman2003,*Mir2007}, the Legget-Garg inequality
\cite{Williams2008,*Palacios-Laloy2010,*Goggin2011}, the quantum
box problem \cite{Resch2004}, and Hardy's paradox \cite{Lundeen2009}.
It has also been used to amplify small experimental effects \cite{Hosten2008,*Dixon2009,*Feizpour2011}
and as feedback for control of a quantum system \cite{Smith2004,*Gillett2010}.
Weak measurements have been demonstrated in both classical \cite{Ritchie1991}
and non-classical systems \cite{Pryde2005}.

The concept of weak measurement is universally applicable to all types
of measurement \cite{Oreshkov2005, Hofmann2010, Dressel2010} but here we introduce
it with a standard model of measurement, the von Neumann model \cite{Von1955}.
In it, a measurement apparatus has a pointer, in an initial position
wavefunction $\left\langle q|\phi_{i}\right\rangle =\phi_{i}(q)\propto\exp\left[-q^{2}/\left(4\sigma^{2}\right)\right]$,
whose momentum $\mathrm{K}$ is coupled with strength $g$ to an system
observable $\mathbf{A}$ via the interaction, $\mathbf{U}=\exp\left(-ig\mathbf{AK}t/\hbar\right)$,
where $t$ is the interaction duration. In a measurement of $\mathbf{A}$,
the position $\mathrm{Q}$ of the pointer is shifted to indicate the
result of the measurement, $\mathrm{A}=a$: $\phi_{i}(q)\rightarrow\phi_{f}(q)=\phi_{i}(q-a)$.
In a standard (i.e.`strong', $gt$ large) measurement, this shift
is much greater than the width $\sigma$ of the pointer and, thus,
unambiguously indicates the measurement result. It also leaves the
system in the associated eigenstate $\left|a\right\rangle $, thereby
radically disturbing it. To perform a weak measurement one reduces
$gt$ such that induced pointer shift is less then $\sigma$, making
the measurement result ambiguous. The benefit is that the system disturbance
is reduced. While a weak measurement on a single system provides little
information, by repeating it on an arbitrarily large ensemble of identical
systems one can determine the \textit{average} measurement result
with arbitrary precision. We call this the `Weak Average' $\left\langle \mathbf{A}^{\mathrm{w}}\right\rangle _{\rho}$.
Unsurprisingly, it is simply equal to the standard quantum expectation
value \cite{Aharonov1990}, $\left\langle \mathbf{A}^{\mathrm{w}}\right\rangle _{\rho}=\mathrm{Tr\left[\mathbf{A}\boldsymbol{\rho}\right]=\left\langle \mathbf{A}^{\mathrm{s}}\right\rangle _{\rho}},$
where the latter indicates $\mathbf{A}$ is measured strongly.

A distinguishing feature of weak measurement is that, in the limit
of zero interaction ($gt=0$), the quantum state of the system remains
unchanged. Subsequent measurements can now provide additional information
about that initial quantum state $\left|\Psi\right\rangle $. Consider
a subsequent strong measurement of observable $\mathbf{C}$ that results
in outcome $c$ (corresponding to eigenstate $\left|c\right\rangle $).
The average result of the weak measurement of $\mathbf{A}$ in the
sub-ensemble of systems giving $\mathrm{C}=c$ is called the `Weak
Value' and is given by \cite{Aharonov1988}, 
\begin{equation}
\left\langle \mathbf{A}^{\mathrm{w}}\right\rangle _{\Psi}^{c}=\frac{\left\langle c\right|\mathbf{A}\left|\Psi\right\rangle }{\left\langle c|\Psi\right\rangle }.\label{eq:weakvalue}
\end{equation}

Surprisingly, the weak value can be outside the range of the eigenvalues
of $\mathbf{A}$ and can even be complex \cite{Lundeen2005,Aharonov1990,Jozsa2007, Hofmann2011}.
Often consideration is limited to its real part, as would be done
in standard measurement \cite{Aharonov1990,Steinberg1995a,Dressel2010}
but the imaginary part also has a physical significance: the evolution
$\mathbf{U}$ not only shifts the average position of pointer but
also the average momentum of the pointer. These two, purely real,
shifts are proportional to the real and imaginary parts of the weak
value, respectively \cite{Lundeen2005,Aharonov1990,Jozsa2007}: $\left\langle \mathbf{A}^{\mathrm{w}}\right\rangle _{\Psi}^{c}=\left\langle \mathbf{Q}\right\rangle _{f}/gt+i\left\langle \mathbf{K}\right\rangle _{f}2\sigma^{2}/gt\hbar$,
where $\left\langle \mathbf{L}\right\rangle _{f}\equiv\left\langle \phi_{f}\right|\mathbf{L}\left|\phi_{f}\right\rangle $.
This result was generalized to other initial pointer wavefunctions
\cite{Mitchison2008}, and discrete pointers (e.g. Qubits or Spins)
\cite{Lundeen2005}. The complex nature of the weak value is what
enables us to directly measure the real and imaginary parts of the
wavefunction and, as we show later, directly measure the Dirac distribution
and density matrix.

\textit{Direct measurement of the quantum wavefunction.}---We now
review our method for the direct measurement of the wavefunction.
The concept is general, however here we consider the case of a discrete
Hilbert space. In this space, one is free to choose the\textit{ }basis
$\left\{ \left\vert a\right\rangle \right\} $ (associated with observable
$\mathbf{A}$) in which the wavefunction will be measured. The method
consists of weakly measuring a projector in this basis $\boldsymbol{\pi}_{a}\equiv\left\vert a\right\rangle \left\langle a\right\vert $,
and post-selecting on a particular value $b_{0}$ of the complementary
observable $\mathbf{B}$. By `complementary' we mean that $\left\langle a|b_{0}\right\rangle =1/\sqrt{N}$
for all $a$, where $N$ is the dimension of the Hilbert space. That
is, the overlap is real and constant as function of $a$. The existence
of state $\left|b_{0}\right\rangle $ is guaranteed by the existence
of at least two mutually unbiased bases (MUB) in any Hilbert space
\cite{Durt2010}. As discussed in the supplementary information of
Ref. \cite{Lundeen2011, Lundeen2006}, the choice of state $b_{0}$ out of the
basis $\left\{ \left\vert b\right\rangle \right\} $ is simply a convention
and is equivalent to choosing a reference frame for the direct measurement,
thereby setting the phases of the basis states in $\left\{ \left\vert a\right\rangle \right\} $.
Using Eq. (\ref{eq:weakvalue}), the quantum state $\left|\Psi\right\rangle $
is given by: $\left\vert \Psi\right\rangle =v\cdot\sum _{a}\left\langle \boldsymbol{\pi}_{a}^{\mathrm{w}}\right\rangle _{\Psi}^{b_{0}}\left\vert a\right\rangle $,
where $\left\langle \boldsymbol{\pi}_{a}^{\mathrm{w}}\right\rangle _{\Psi}^{b_{0}}$
is the weak value and $v$ is a constant that is independent of $a$.
Thus, by stepping through the values of $a$ in a series of weak measurements
one can directly measure $\left\vert \Psi\right\rangle $ represented
in the $a$ basis.

\textit{Weak measurement of mixed states.}---The weak value of a system
described by a density operator $\rho$ was first considered in \cite{Wiseman2002}
and shown to be: 
\begin{equation}
\left\langle \mathbf{A}^{\mathrm{w}}\right\rangle _{\rho}^{c}=\frac{\left\langle c\right|\mathbf{A}\boldsymbol{\rho}\left|c\right\rangle }{\left\langle c\right|\boldsymbol{\rho}\left|c\right\rangle }.\label{eq:wv_mixed}
\end{equation}
Applying this to our direct measurement method we find, $\left\langle \boldsymbol{\pi}_{a}^{\mathrm{w}}\right\rangle _{\rho}^{b_{0}}=\left\langle b_{0}|a\right\rangle \left\langle a\left|\boldsymbol{\rho}\right|b_{0}\right\rangle /\left\langle b_{0}\right|\boldsymbol{\rho}\left|b_{0}\right\rangle $.
Only $2N$ real parameters are found by scanning $a$. This will not
generally be sufficient to determine all the parameters in $\rho$,
which has $N^{2}-1$ real parameters. Consequently, our method for
the direct measurement of the wavefunction cannot be used to determine
a mixed state.

\textit{Direct measurement of the Dirac distribution.}---We now consider
what happens if one replaces the strong measurement of $\mathbf{B}$
with a weak measurement. Specifically, we investigate the weak measurement
of the product of projectors from the two MUB, $\mathbf{S}_{ab}\equiv\bigl|b\bigr\rangle\bigl\langle b|a\bigr\rangle\bigl\langle a\bigr|$
with no post-selection whatsoever. We wish to measure its weak average.
Although, a non-Hermitian operator $\mathbf{A}$ is typically considered
to be unobservable, later, we shall outline specific methods to \textit{weakly}
measure it. Surprisingly, we show that even if $\mathbf{A}$ is non-Hermitian
$\left\langle \mathbf{A}^{\mathrm{w}}\right\rangle _{\rho}=\mathrm{Tr\left[\mathbf{A}\boldsymbol{\rho}\right]}$
still holds for the weak average. In this case, $\left\langle \mathbf{A}^{\mathrm{w}}\right\rangle _{\rho}$
is complex with a physical significance similar to that of the weak
value (i.e. shifts in the position and momentum of the pointer). For now, we use this result to find the weak average of $\mathbf{\mathbf{S}}_{ab}$:

\begin{equation}
\left\langle \mathbf{S}_{ab}^{\mathrm{w}}\right\rangle _{\rho}=\mathrm{Tr}\left[\mathbf{S}_{ab}\boldsymbol{\rho}\right]=\left\langle a\right|\boldsymbol{\rho}\left|b\right\rangle \left\langle b|a\right\rangle =S_{\rho}(a,b),\label{eq:jointweak-1}
\end{equation}
where $S_{\rho}(a,b)$ is the discrete Hilbert space version of the
Dirac distribution as defined in \cite{Chaturvedi2006}. Dirac introduced
this phase-space distribution as a way to represent a quantum operator
$\mathbf{O}$ in his 1945 paper \cite{Dirac1945}, {}``On the Analogy
Between Classical and Quantum Mechanics.'' In various guises it has been investigated periodically during last half-century \cite{Johansen2007}. In optics, variations
of the Dirac distribution have been used widely, appearing in Walther's
definition of the radiance function in radiometry \cite{Walther1968}
and Wolf's specific intensity \cite{Wolf1976} (as pointed out in
\cite{Chaturvedi2006}). If $\mathbf{O}=\boldsymbol{\rho}$, the Dirac
distribution is a representation of the quantum state of a system.
For instance, the joint weak measurement of a position $x$ and a
momentum $p$ (i.e. $\mathbf{S}_{xp}\equiv\bigl|p\bigr\rangle\bigl\langle p|x\bigr\rangle\bigl\langle x\bigr|$)
on a mixed state $\boldsymbol{\rho}$ gives the phase-space version
of the Dirac distribution, $S_{\rho}(x,p)$, which, although it is
complex, shares many of the desired features of a quasi-probability
distribution \cite{Chaturvedi2006}.

In our weak measurement, if one scans $a$ and $b$, so as to directly
measure the Dirac distribution over all values of $(a,b)$, one completely
determines the density operator. But in order to actually calculate
the density operator from the Dirac distribution one must know $\left\langle b|a\right\rangle =\exp\left(i\theta_{ab}\right)/\sqrt{N}$.
Since it is not generally known what are the bases in the MUB set
(for any given Hilbert space) a general formula for $\theta_{ab}$
is also unknown. However, if $\left\{ \left\vert a\right\rangle \right\} $
is taken to be the standard basis (i.e. $\sum_{a=0}^{N}\left|a\right\rangle \left\langle a\right|=\mathbf{I},$
the identity operator) then one MUB, which we take to be $\left\{ \left\vert b\right\rangle \right\} $,
will always be the Fourier basis \cite{Bengtsson2007,Durt2010}, $\left\vert b\right\rangle =\sum_{a=0}^{N-1}\left\vert a\right\rangle \exp\left(i2\pi ab/N\right)/\sqrt{N}$.
In this case, $\theta_{ab}=-2\pi ab/N$, where $a$ and $b$ are integers
solely used to enumerate the states such that $0\leq a,b\leq N-1$.
With these choices for our complementary bases the density operator
is simply related to the Dirac distribution by a Discrete Fourier
Transform, $\rho_{a_{1}a_{2}}=\sum_{b=0}^{N-1}S_{\rho}\left(a_{1},b\right)e{}^{i2\pi b(a_{1}-a_{2})/N},$
where $\rho_{a_{1}a_{2}}=\left\langle a_{1}\left|\boldsymbol{\rho}\right|a_{2}\right\rangle $.
This explicitly shows the weak average, $\left\langle \mathbf{S}_{ab}^{\mathrm{w}}\right\rangle _{\rho}$,
contains the same information as the density operator. Much like the
Wigner function, the Dirac distribution can used to find the expectation
value of an observable through a simple overlap integral \cite{Chaturvedi2006}. Unlike the Wigner function, it is compatible with Bayes' law and, thus, is consistent with a quantum analog of classical determinism \cite{
Hofmann2011a}.

\textit{Direct measurement of the density operator}.---While quasi-probability
distributions are informationally equivalent to the density operator,
they are less commonly known and used. Motivated by a desire to understand
the nature of the density operator we now describe how to measure
it directly in a given basis. Consider the weak measurement of the
product of three projectors, $\boldsymbol{\Pi}_{a_{1}a_{2}}=\boldsymbol{\pi}_{a_{2}}\boldsymbol{\pi}_{b_{0}}\boldsymbol{\pi}_{a_{1}}$,
where $\boldsymbol{\pi}_{a_{1}}=\left\vert a_{1}\right\rangle \left\langle a_{1}\right\vert $,$\boldsymbol{\pi}_{b_{0}}=\left\vert b_{0}\right\rangle \left\langle b_{0}\right\vert ,$
and $\boldsymbol{\pi}_{a_{2}}=\left\vert a_{2}\right\rangle \left\langle a_{2}\right\vert $
and $b_{0}$ is chosen so that $\left\langle a|b_{0}\right\rangle =1/\sqrt{N}$
for all $a$ (but is not required to be from the Fourier basis). As
before, there is no post-selection. The weak average is: 
\begin{equation}
\left\langle \boldsymbol{\Pi}_{a_{1}a_{2}}^{\mathrm{w}}\right\rangle _{\rho}=\left\langle a_{1}\left|\boldsymbol{\rho}\right|a_{2}\right\rangle /N=\rho_{a_{1}a_{2}}/N.\label{eq:density}
\end{equation}
Thus, each element $\rho_{a_{1}a_{2}}$ of the density matrix in any
chosen basis (here $\left\{ \left\vert a\right\rangle \right\} $)
can be measured directly by weakly measuring the corresponding projectors,
$\boldsymbol{\pi}_{a_{1}}$ and $\boldsymbol{\pi}_{a_{2}}$, in between
which is a third weak measurement of $\boldsymbol{\pi}_{b_{0}}$.
The proportionality constant $N^{-1}$ can be eliminated through the
normalization of the density matrix so that $\mathbf{\mathrm{Tr}}\left[\boldsymbol{\rho}\right]=1$.
Keeping $b_{0}$ fixed while scanning $a_{1}$ and $a_{2}$ allows
one to map out the entire density matrix.

\textit{Weak measurement of products of complementary variables.}---One
cause for concern in our two direct measurement methods is that $\mathbf{S}_{ab}$
and $\boldsymbol{\Pi}_{a_{1}a_{2}}$ are not Hermitian, which, according
to the postulates of quantum mechanics, means they are not observable
\cite{Shankar1994}. Indeed, coupling such operators to a pointer
via the von Neumann interaction (as in $\mathbf{U}$) leads to an
unphysical non-unitary evolution. This issue can be circumvented by
dividing the measurement into a sequence of unitary von Neumann interactions.
Each has a pointer beginning in same initial state $\phi_{i}(q)$.
We now describe a pair of schemes that use this strategy to weakly
measure the product of two non-commuting observables $\mathbf{E}$
and $\mathbf{F}$ thereby measuring their weak average $\left\langle \left(\mathbf{EF}\right)^{\mathrm{w}}\right\rangle _{\rho}$.
In the process, we will show that the weak average $\left\langle \left(\mathbf{EF}\right)^{\mathrm{w}}\right\rangle _{\rho}=\mathrm{Tr}\left[\mathbf{EF}\boldsymbol{\rho}\right]$.
And later we will show that weakly measuring just two observables
is sufficient to implement both direct measurement methods.

\textit{Scheme 1}: The first scheme follows a commonly used strategy
for standard (strong) measurements: perform independent measurements
of two observables and correlate the results to find the observables'
product. With von Neumann measurements the total evolution is $\mathbf{U}_{T}\equiv\exp\left(ig_{2}\mathbf{E}\mathbf{K}_{2}t/\hbar\right)\exp\left(ig_{1}\mathbf{F}\mathbf{K}_{1}t/\hbar\right)$,
where the subscripts indicate observables on pointer 1 or 2. If $\mathbf{E}$
and $\mathbf{F}$ commute, strong measurements give $\left\langle \left(\mathbf{EF}\right)^{\mathrm{s}}\right\rangle _{\rho}\propto\left\langle \mathbf{Q}_{1}\mathbf{Q}_{2}\right\rangle _{f}$.
The weak measurement version of this strategy was proposed in \cite{Resch2004a}.
It was simplified in \cite{Lundeen2005} by forming the composite
operator $\mathfrak{\boldsymbol{a}}\equiv\mathbf{Q}/2\sigma+i\mathbf{K}\sigma/\hbar$,
which has the form of an annihilation operator (i.e. $\mathfrak{\boldsymbol{a}}\left|\phi_{i}\right\rangle =0$),
so that the standard weak value (Eq. \ref{eq:weakvalue}) has the
simple form $\left\langle \mathbf{A}^{\mathrm{w}}\right\rangle _{\Psi}^{c}=\left(2\sigma/gt\right)\left\langle \mathfrak{\boldsymbol{a}}\right\rangle _{f}$.
Following \cite{Lundeen2005}, one can show that in the limit $g_{1}g_{2}\left(t/\sigma\right)^{2}\ll1$,
the evolution $\mathbf{U}_{T}$ induces the pointer shifts $\left\langle \left(\mathbf{EF}\right)^{\mathrm{w}}\right\rangle _{\rho}=\mathrm{Tr}\left[\mathbf{EF}\boldsymbol{\rho}\right]=g_{1}g_{2}\left(2\sigma/t\right)^{2}\left\langle \mathfrak{\boldsymbol{a}}_{1}\mathfrak{\boldsymbol{a}}_{2}\right\rangle _{f}$.
This scheme was demonstrated experimentally in \cite{Lundeen2009}
for products of commuting observables. Ref. \cite{Mitchison2008}
showed that it is valid even for non-commuting observables $\mathbf{E}$
and $\mathbf{F}$ if they are measured sequentially, as in $\mathbf{U}_{T}$.
(This result can be generalized to an $n$-product observable, such
as the triple product $\boldsymbol{\Pi}_{a_{1}a_{2}}$ \cite{Lundeen2005,Mitchison2008}.)
Thus, just as with strong measurement, by performing independent measurements
of each observable and then evaluating a joint expectation value on
the pointers one can measure $\left\langle \left(\mathbf{EF}\right)^{\mathrm{w}}\right\rangle _{\rho}$.

\textit{Scheme 2}: The second scheme measures $\mathbf{F}$ and then,
conditioned on the result, measures $\mathbf{E}$, thereby measuring
their product. A von Neumann interaction couples $\mathbf{F}$ to
a first pointer, which shifts the pointer's position. With a strength
proportional to this shift a second von Neumann interaction couples
$\mathbf{E}$ to a second pointer. This conditional sequential measurement
is described by the total interaction $\mathbf{U}_{\mathrm{D}}\equiv\exp\left(-ig_{2}\mathbf{E}\mathbf{K}_{2}\mathbf{Q}_{1}t/\hbar\right)\exp\left(-ig_{\mathrm{D}}\mathbf{F}\mathbf{D}_{1}t/\hbar\right)$,
where the rightmost interaction couples to either $\mathbf{D}=\mathbf{Q}$
or $\mathbf{K}$, and the subscripts refer to pointer 1 or 2. For
$\mathbf{D}=\mathbf{K}$ our weak measurement of $\mathbf{F}$ is
a standard von Neumann interaction. In the limit of $g_{K}g_{2}t^{2}/\sigma\ll1$,
the evolution $\mathbf{U}_{\mathrm{K}}$ shifts the position of pointer
2 by $\left\langle \mathbf{Q}_{2}\right\rangle _{f}=\left(g_{K}g_{2}t^{2}\right)\mathrm{Re}\left\{ \mathrm{Tr}\left[\mathbf{EF}\boldsymbol{\rho}\right]\right\} $.
However, $\left\langle \mathbf{K}_{2}\right\rangle _{f}=0$, leaving
us without $\mathrm{Im}\left\{ \mathrm{Tr}\left[\mathbf{EF}\boldsymbol{\rho}\right]\right\} $.
The imaginary component can be found by coupling $\mathbf{F}$ to
the position, rather than momentum, of the first pointer. If we set
$\mathbf{D}=\mathbf{Q}$, then in the limit $g_{Q}g_{2}t^{2}\sigma/\hbar\ll1$,
the evolution $\mathbf{U}_{\mathrm{Q}}$ shifts the momentum of the
second pointer by $\left\langle \mathbf{Q}_{2}\right\rangle _{f}=\left(2g_{\mathrm{Q}}g_{2}t^{2}\sigma^{2}/\hbar\right)\mathrm{Im}\left\{ \mathrm{Tr}\left[\mathbf{EF}\boldsymbol{\rho}\right]\right\} $.
In summary, conditional two sequential weak measurements allow us
to measure the real and imaginary components of expectation value
of a product operator.

Both of the schemes can be followed by post-selection of some other
observable and, in that case, they would give the weak value of $\mathbf{EF}$,
which can be complex. However, our schemes show that, even without
post-selection, the weak average of a non-Hermitian $\mathbf{EF}$
will be complex. This may come as a surprise since post-selection
is often cited as the mechanism for anomalous weak values \cite{Aharonov1990,Ritchie1991,Dixon2009}.

\textit{How to substitute a strong measurement for a weak measurement
of one observable in a product.}---We now show that the weak measurement
of $\mathbf{CG}$ is equivalent to first weakly measuring $\mathbf{G}$
followed by a strong measurement of $\mathbf{C}$, $\left\langle \left(\mathbf{CG}\right)^{\mathrm{w}}\right\rangle =\left\langle \mathbf{C}^{\mathrm{s}}\mathbf{G}^{\mathrm{w}}\right\rangle $.
This can greatly simplify both of our proposed methods. In analogy
to a standard joint expecation value, by $\left\langle \mathbf{C}^{\mathrm{s}}\mathbf{G}^{\mathrm{w}}\right\rangle $
we mean that for each measured outcome of the strong measurement of
$\mathbf{C}$ one multiplies the corresponding eigenvalue $c$ by
its probability $P(c)$ by the weak value $\left\langle \mathbf{G}^{\mathrm{w}}\right\rangle _{\rho}^{c}$
to find the average, $\sum_{c}cP(c)\left\langle \mathbf{G}^{\mathrm{w}}\right\rangle _{\rho}^{c}\equiv\left\langle \mathbf{C}^{\mathrm{s}}\mathbf{G}^{\mathrm{w}}\right\rangle $.
Using Eq. (\ref{eq:wv_mixed}), it follows that, $\left\langle \mathbf{C}^{\mathrm{s}}\mathbf{G}^{\mathrm{w}}\right\rangle =\sum_{c}c\left\langle c\right|\mathbf{G}\boldsymbol{\rho}\left|c\right\rangle =\mathrm{Tr}\left[\mathbf{CG}\boldsymbol{\rho}\right]=\left\langle \left(\mathbf{CG}\right)^{\mathrm{w}}\right\rangle $.
In other words, a joint weak-strong measurement of $\mathbf{G}$ and
$\mathbf{C}$, respectively, will have the average result, $\mathrm{Tr}\left[\mathbf{CG}\boldsymbol{\rho}\right]$.
Note that because both the weak value and corresponding the pointer
expectation values (e.g. $\left\langle \mathbf{Q}\right\rangle _{f}$)
are normalized by $P\left(c\right)$, in an actual experiment the
pointer signal will be proportional to $\left\langle c\right|\mathbf{G}\boldsymbol{\rho}\left|c\right\rangle $
directly, removing the need to find the weak value.

Consequently, we can directly measure the full Dirac distribution
and density matrix by measuring the correlations between the weak
measurement and the subsequent strong measurement outcome while we
scan the weak measurement. For the density matrix one would weakly
measure $\boldsymbol{\pi}_{b_{0}}\boldsymbol{\pi}_{a_{1}}$ using
scheme 1 or 2 and then strongly measure of $\boldsymbol{\pi}_{a_{2}}$.
Alternatively, one can instead strongly measure $\mathbf{A}$, which
projects on the eigenstates $\left\{ \boldsymbol{\pi}_{a}\right\} $
in parallel. This has the advantage that the final measurement no
longer must be scanned. Similarly, in the experiment in \cite{Lundeen2011},
replacing the slit with a camera (with a preceding polarization analyzer)
in the momentum plane is the only modification necessary to directly
measure the Dirac distribution for the transverse density matrix of
a photon. Evidently, dividing the joint measurement into weak and
strong parts simplifies each of the two new direct measurements considered
in this paper (given in Eq. (\ref{eq:jointweak-1}) and Eq. (\ref{eq:density}))
and makes them feasible with existing technology (e.g. weak measurement
of the product of two operators \cite{Lundeen2009}). 

\textit{Conclusion}.---In this work, we have shown that by weakly
measuring pairs or triple products of complementary variables of a
system it is possible not only to directly measure its wavefunction
but, also, its density operator. To measure over the extent of wavefunction
we only need to scan the first variable of a complementary pair. To
determine the density operator through its Dirac distribution we must,
additionally, scan the second variable. To determine the density matrix
directly, one weakly measures the product of a variable, its complementary
variable, and the first variable again. Leaving the complementary
variable fixed at one value while rastering the values of the other
two completely maps the density matrix, one element at a time. This
procedure thus provides the density matrix with an operational definition,
it is the average result of a joint weak measurement of a variable,
then its complementary variable, then the original variable.

These methods also provide alternatives to standard state tomography
that have three key advantages: One, they are simpler in that they
only require measurements in two of the system's bases. Two, they
do not require a global reconstruction --- states can be determined
locally, point by point. And three, the amount of state disturbance
can be minimized. Thus, in principle, we can characterize quantum
states in situ, for instance, in the middle of quantum computation
circuits, or during chemical reactions, without disturbing the process
in which they feature.

\begin{acknowledgments}
We thank Aephraim Steinberg for useful discussions.
\end{acknowledgments}

\bibliography{gen_wavefunction_prl}

\end{document}